\documentclass[review]{elsarticle}

\usepackage{lineno,hyperref}
\usepackage{stfloats}
\usepackage{bm}
\usepackage{upgreek}
\usepackage{amsmath}
\usepackage{amsfonts}

\journal{arXiv}

\begin{document}

\begin{frontmatter}

\title{Quantitative imaging for complex-objects via a single-pixel detector}


\author[addressa]{Xianye Li}

\author[addressa]{Yafei Sun}

\author[addressa]{Yikang He}

\author[addressb]{Xun Li}

\author[addressa]{Baoqing Sun\corref{mycorrespondingauthor}}
\cortext[mycorrespondingauthor]{Corresponding author}
\ead{baoqing.sun@sdu.edu.cn}

\address[addressa]{School of Information Science and Engineering, Shandong University, Qingdao, 266237, China}
\address[addressb]{Department of Electrical and Computer Engineering, McMaster University, Hamilton, ON L8S 4k1, Canada}

\begin{abstract}
    Quantitative phase imaging (QPI) is important in many applications such as microscopy and crystallography. To quantitatively reveal phase information, people could either employ interference to map phase distribution into intensity fringes, or analyze intensity-only diffraction patterns through phase retrieval algorithms. Traditionally, both of these two ways use pixelated detectors. In this work, a novel QPI scheme is reported inspired by single-pixel camera (SPC), which adopts the principle of SPC that retrieves images through structured illumination and corresponding single-pixel signals. Particularly for complex-valued imaging, the structured illumination is performed in the phase domain, and a point detector with restricted sensor size detects the intensity of zero-frequency area. Based on the illumination structures and point signals, a complex image is reconstructed by running a phase retrieval algorithm. This approach is universal for various wavelengths, and needs no a priori information of the targets. Both simulation and experiment show that our single-pixel QPI scheme exhibits great performance even with objects in an extremely rough phase distribution.  
\end{abstract}


\end{frontmatter}

\section{Introduction}

Phase of light carries important information but is naturally missed in normal imaging process, as photodetectors from the human retina to modern digital cameras only respond to power but not phase of the field. To extend our vision capacity to this unrevealed domain, various phase imaging technologies have been developed\cite{QPIreview,TIE,phasecontrast}. Among them, a majority category of approaches is based on interferometry. The importance of interference in the revealing phase was realized as early as Abbe \cite{abbe}, when he described a microscope image as the interference effect of a diffraction phenomenon. Based on this idea, Zernike designed his phase contrast microscope\cite{zernike1955discovered}. 
Subsequently, by combining the principle of interference with holography, quantitative phase imaging (QPI) is developed that enables quantitative measurement of complex objects \cite{Digitalholography,Phase-shifting}. In these days, interferometric QPI technique has been wildly used as a powerful imaging or measurement method in many fields, including wavefront sensing, biomedical imaging, 3D measurement and so on \cite{wavefrontsensing,QPIBio,holo3D}. Nevertheless, it needs to be emphasized that interferometric QPI has a high standard requirement for the illumination coherence, and meanwhile the interferometric arrangement always increases the apparatus complexity. Both of these two limitations have prevented its further applications. 

When interferometry is inapplicable, to retrieve phase from diffraction intensity patterns becomes a classical phase retrieval problem. It originates in the field of X-ray crystallography, where periodic structure of crystalline media is estimated by recovering the phase information missed in coherent diffraction patterns. Related technique is therefore named as coherent diffraction imaging (CDI)\cite{wolf,phasereview}. Due to its diffraction imaging mechanism, CDI can be easily achieved in a lensless profile and therefore appropriate for short-wavelength spectrum such as X-ray, electrons and neutrons\cite{cdixray,electron,neutron}. Subsequently, the achievements in efficient phase retrieval algorithms enable CDI to be a powerful QPI method for general complex-amplitude objects \cite{GS, fienup}. On the other hand, however, phase retrieval is a NP-hard problem, conventional algorithms usually suffer from inherent ambiguities \cite{nphard}. The application of CDI therefore always requires some extra constraints, e.g., a clear edge or relative small phase variance of the object \cite{support}. 
Additionally, the extremely uneven diffraction spectrum of natural images has a very high requirement for the detector dynamics range, which in turn increases measurement difficulties \cite{Fourier}. Coherent modulation imaging \cite{CMI1,CMI2} tries to overcome these limitations by introducing  a downstream modulator. However, The function of the modulator and the geometric setup parameters are the a priori information, which needs to be carefully calibrated or characterized. This renders the imaging process more complex and less robust.  Therefore, to realize a robust and applicable phase imaging scheme for more universal applications is always of great demand.

Single-pixel imaging (SPI) is a novel imaging method inspired by both ghost imaging and compressive sensing\cite{SPIreview, abedi2020single, SPICS}. It gets intensively studied for its minimized requirement on detector resolution and great application potential at spectral wavelength. Therefore, many researchers focus on their works on reconstructing the phase information through SPI. However, most of these works are still based on the traditional holography or phase-shifting strategies \cite{csholo, liu2019complex, liu2018single, hu2019single}, which still require additional reference arm for interferometry.

In this work, we develop an efficient and robust diffraction imaging (DI) scheme based on SPI. Our single-pixel DI (SPDI) scheme is configured by introducing phase modulation in a single-pixel camera with coherent illumination. An area-restricted detector is placed in the far-field to detect the $0$th order intensity of the diffraction spectrum. Finally, we employ a truncated amplitude flow (TAF) algorithm for reconstruction. Besides the establishment of a theoretical framework, we have also carried out experiments to demonstrate its feasibility. These experiments show that our SPDI scheme is capable to perform high quality complex-valued reconstruction even for some rigorous objects.

\section{Theory and method}
Conventional SPI retrieves intensity-only images via correlation measurement between structured illumination and non-pixelated detection. Illuminations with different intensity modulations are projected onto an object. Under each modulation, the total light power after the object is collected by a single-pixel detector. The value can be written as
\begin{equation}
  \begin{aligned}
    y_{i} =\sum_{u=1}^{m}\sum_{v=1}^{n}\bm{P}_{i}\left ( u,v \right ) \cdot \bm{T}\left ( u,v \right )   \;\;\;\; i = 1, 2, \cdots , M
  \end{aligned}
\end{equation}
where ($u, v$) are coordinates in the image plane, $i$ denotes the $i$th measurement, $M$ denotes the total number of measurements, and "$\cdot$" denotes the Hadamard product. $\bm{P}_i (u, v)  \in \mathbb{R}^{m \times n}$ is the intensity distribution of the $i$th modulation, and $\bm{T}(u, v) \in \mathbb{R}^{m \times n} $ is the transmittance or reflectance function of the target, $m\times n$ is the imaging resolution determined by the illumination. An image can be reconstructed from the correlation between $y_{i}$ and $\bm{P}_i (u,v)$:
\begin{equation}
\hat{\bm{T}}\left ( u,v\right )=\frac{1}{M}\sum_{i=1}^{M}\left (y_{i}-\left \langle y\right \rangle\right )\bm{P}_{i}\left ( u,v\right )
\label{equation2}
\end{equation}
where "$\left \langle \cdot \right \rangle$" stands for an ensemble average of $M$ measurements. $\hat{\bm{T}}\left ( u,v\right )$ is the estimation of target $\bm{T}\left ( u,v\right )$, which gradually approaches $\bm{T}\left ( u,v\right )$ as $M$ increases.

In this work, we establish a SPDI scheme by extending SPI to the complex domain. Different from intensity SPI, structured illumination in our scheme is generated by performing phase modulation over a coherent light field. In the detection end, as is widely discussed in ghost imaging, a pinhole detector is mandatorily used to collect phase-sensitive signals in the far field. For a complex object $ \bm{T}\left ( u,v \right ) \in \mathbb{C}^{m \times n}$, the far-field diffraction pattern in discrete coordinate can be represented as
\begin{small}
\begin{equation}
\bm{I}\left ( \xi ,\eta  \right )=\left |\sum_{u=1}^{m}\sum_{v=1}^{n} \bm{P}_{i}\left ( u,v \right ) \cdot \bm{T}\left ( u,v \right ) \cdot exp[-j2\pi (u \xi/m+v \eta/n)]\right |^{2}
\label{farfield}
\end{equation}
\end{small}
where $\bm{P}_{i} \in \mathbb{C}^{m \times n}$ is the $ith$ complex-valued modulation pattern under coherent illumination, "$ \left | \cdot  \right |^{2} $" represents intensity measurement of the complex filed, and ($\xi ,\eta$) are coordinates of the far field. Here the constant coefficient is omitted in far field calculation for brief expression.
In the far field, the pinhole detector measures the intensity of the $0$th order spectrum  as $\bm{I}\left ( 0, 0  \right )$, therefore, the detection value is
\begin{equation}
  \begin{aligned}
  y_{i} =\left |\sum_{u=1}^{m}\sum_{v=1}^{n}\bm{P}_{i}\left ( u,v \right ) \cdot \bm{T}\left ( u,v \right ) \right |^{2}
  \end{aligned}
  \label{fourier0}
\end{equation}

To omit the integrating processes in Eq.~(\ref{fourier0}), we reshape the pattern distribution $\bm{P}_{i}$ and object distribution $\bm{T}$ to two vectors and rewrite the detection process as 
\begin{equation}
  \begin{aligned}
  y_{i} =\left |   \bm{a}_{i}^{T}\bm{x}  \right |^{2}
  \end{aligned}
  \label{det}
\end{equation}
where the vectors $\bm{a}_{i} \in \mathbb{C}^{N}$ and $\bm{x} \in \mathbb{C}^{N}$ are one-dimensional representations of $\bm{P}_{i}$ and $\bm{T}$, $N$ represents the imaging resolution, which is equals to $m \times n$.

\begin{figure}[t!]
  \centering
  \includegraphics[width=0.7\linewidth]{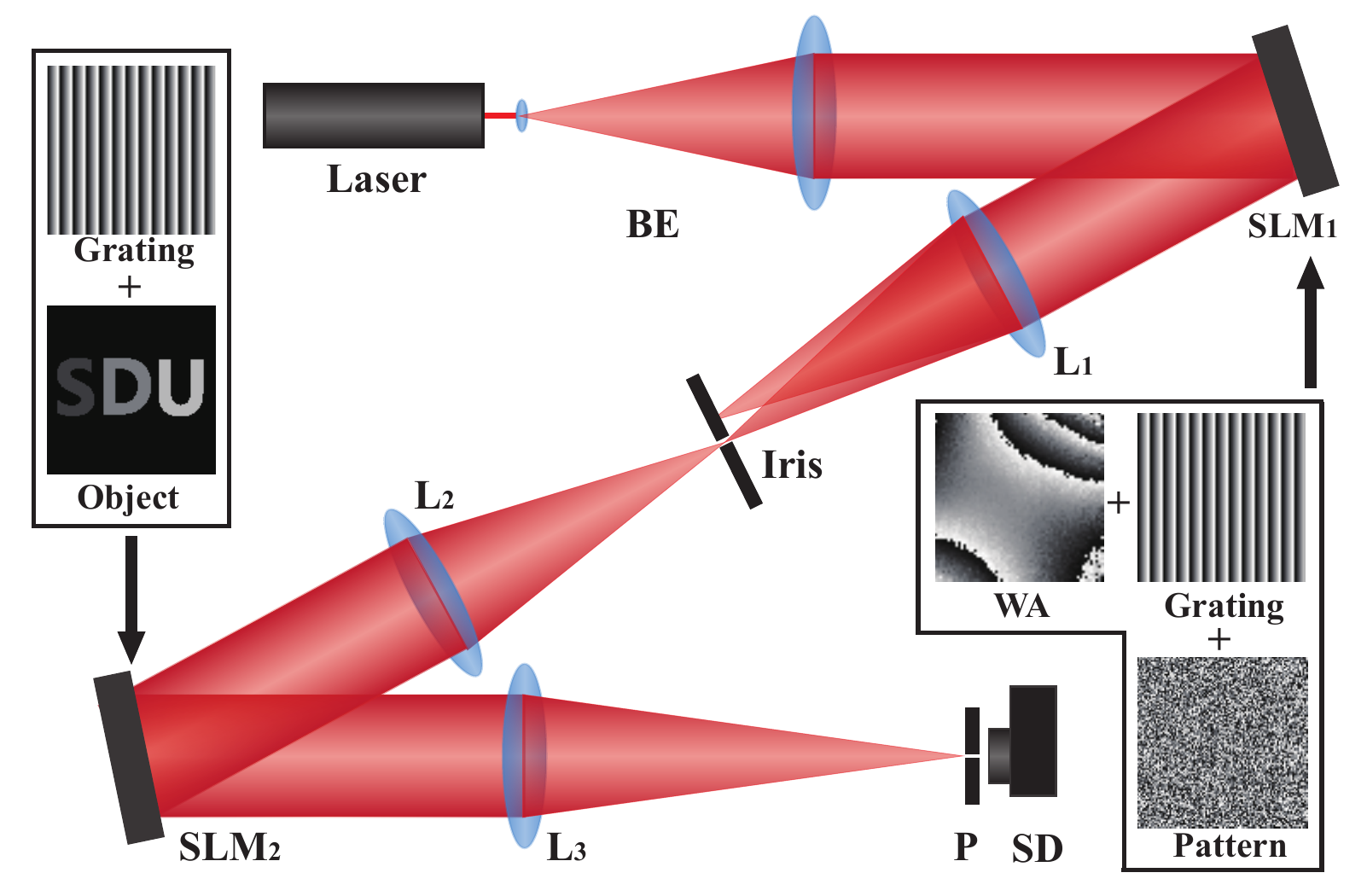}
  \caption{Schematic diagram of the SPDI scheme. A laser beam is expanded by a beam expander (BE) and then illuminates a SLM ($ \rm SLM_{1}$) for phase modulation. Modulated light passes through a $4$-$f$ system and is incident onto the object (represented by $ \rm SLM_{2}$). ${\rm L_{3}}$ is a Fourier transform lens, its front focal plane overlapping the object plane. On the back focal plane a single-pixel detector (SD) is placed for signal collection. P: Pinhole; WA: wavefront aberration.}
  \label{fig1}
\end{figure}

Until now we have established a framework of a SPDI experiment that can be represented as Eq.~(\ref{det}). To solve complex unknowns from Eq.~(\ref{det}) is a typical phase retrieval process. It can be regarded as an iterative estimation process by minimizing an empirical loss function. Adopting the least-square criterion, we define our loss function $l(\hat{\bm{x}})$ as the residual between the detected and estimated amplitudes of zero-frequency point. Then the calculation process can be represented as 
\begin{equation}
  \textup{minimize}\;\;\;\; l(\hat{\bm{x}})=\frac{1}{2M}\sum_{i=1}^{M}(b_{i}- \left | \bm{a}_{i}^{T}\hat{\bm{x}} \right |)^{2}
  \label{loss}
\end{equation}
 where $b_{i}=\sqrt{y_{i}} $. However, as mentioned above, it is a NP-hard problem to minimize such a non-convex quadratic function from intensity-only detections. 
To solve this ill-posed problem, one way is to introduce additional constraints on the known, e.g., amplitude constraint or sparsity condition \cite{phasereview, phasecodecpr}. Another approach is to perform redundant sampling \cite{coded}.

In order to make an universal approach, we apply redundant sampling strategy in our scheme. Redundant sampling means to acquire more samples than the size of imaging resolution \cite{oversampling}. It is widely used in phase retrieval problem as it can provide more independent equations to determine the unique solution of non-linear detection equations. In our SPI scheme, redundant samplings can be achieved through dynamic mask modulation, and the advances of high performance spatial light modulators (SLM) make it convenient for the experimental implementation. Finally, solving Eq.~(\ref{loss}) from redundant data could have different approaches. For examples, semi-definite programming \cite{SDP,phasemax} and non-convex optimization \cite{TAF,Wirtingerflow} are two classical schemes. In our work, we tried out various algorithms with both simulation and Experimental data. Based on the comparison of different reconstruction results, we adopt truncated amplitude flow (TAF) algorithm for its performance both in recovering quality and calculation speed. TAF algorithm is an outstanding non-convex optimization algorithm that apply a generalized gradient decent method to minimize the loss function in a phase problem\cite{TAF}.
 In our work, TAF algorithm can work out reasonable reconstructions with various generalized objects, even for those with steep phase changes. Previous work has reported that a $400\%$ sampling ratio is the minimum requirement for determining the unique solution of a complex object \cite{signalrec,ratio}. In our scheme, we will show that a $200\%$ sampling ratio is able to generate a clearly reconstruction. If we take into account that each pixel of the reconstruction contains two variables, this is almost a reconstruction in a full sampling condition.

\section{Experimental design}

\subsection{Experimental arrangement}

Details of our Experimental arrangement is illustrated in Fig.~\ref{fig1}. Phase modulation is performed by  illuminating a SLM ($ \rm SLM_{1}$) with a collimated quasi-monochromatic light source. Light field carrying modulation information is then projected onto an object (represented as $ \rm SLM_{2}$) through a $4$-$f$ system. Behind the object, a Fourier transform lens ($ \rm L_{3}$, $f=100$mm) is placed to transform the wavefront at the object plane to its frequency plane, where a point detector is placed for signal collection.


In the experiment, the quasi-monochromatic illumination source is consisted of a He-Ne laser (Thorlabs HNL150LB) at a wavelength of $\lambda = 632.8$nm and a beam expander. To maximize modulation efficiency, blazed grating holograms are implemented on both of the two SLMs to separate direct-reflected light away from effective modulation. 
To do this, an iris is used in the $4$-$f$ system to filter out all but the first-order diffraction spectrum. At the detection end, a pinhole ($ \rm P $) with a diameter of 25$\upmu$m  is placed in front of the amplified photodetector (Thorlabs PDA-100A). The detection area is therefore restricted to ensure that only the $0$th order component within the effective diffraction spectrum of the object is detected. It should be emphasized that, besides the SLM-based objects, we also image two transmissive objects, in which case the detection part is in a transmissive layout along the $4$-$f$ system.

Before imaging any specific objects, wavefront aberration of the system should be determined. This aberration could be caused by the phase unevenness of the illumination source as well as the SLMs and imaging lenses. The wavefront measured without any objects can be seen as the systematic aberration of the proposed system. Once this systematic aberration (also illustrated in Fig.~\ref{fig1}) is measured, a corresponding wavefront correction will be taken into account by subtracting this aberration from each projected pattern to avoid unnecessary defection.
Finally, for $ \rm SLM_{1}$ (Holoeye Pluto-VIS-016), an area of $800\times800$ pixels is used for spatial modulation. By binning each $8\times8$ pixels into a super-pixel, the actual reconstruction resolution is $100\times 100$ pixels. Unless stated otherwise, all the reconstructions are done with a sampling ratio of $600\%$. That is, each image is reconstructed from $60,000$ samplings.

\subsection{Modulation design}

\begin{figure*}[t!]
  \centering
  \includegraphics[width=\linewidth]{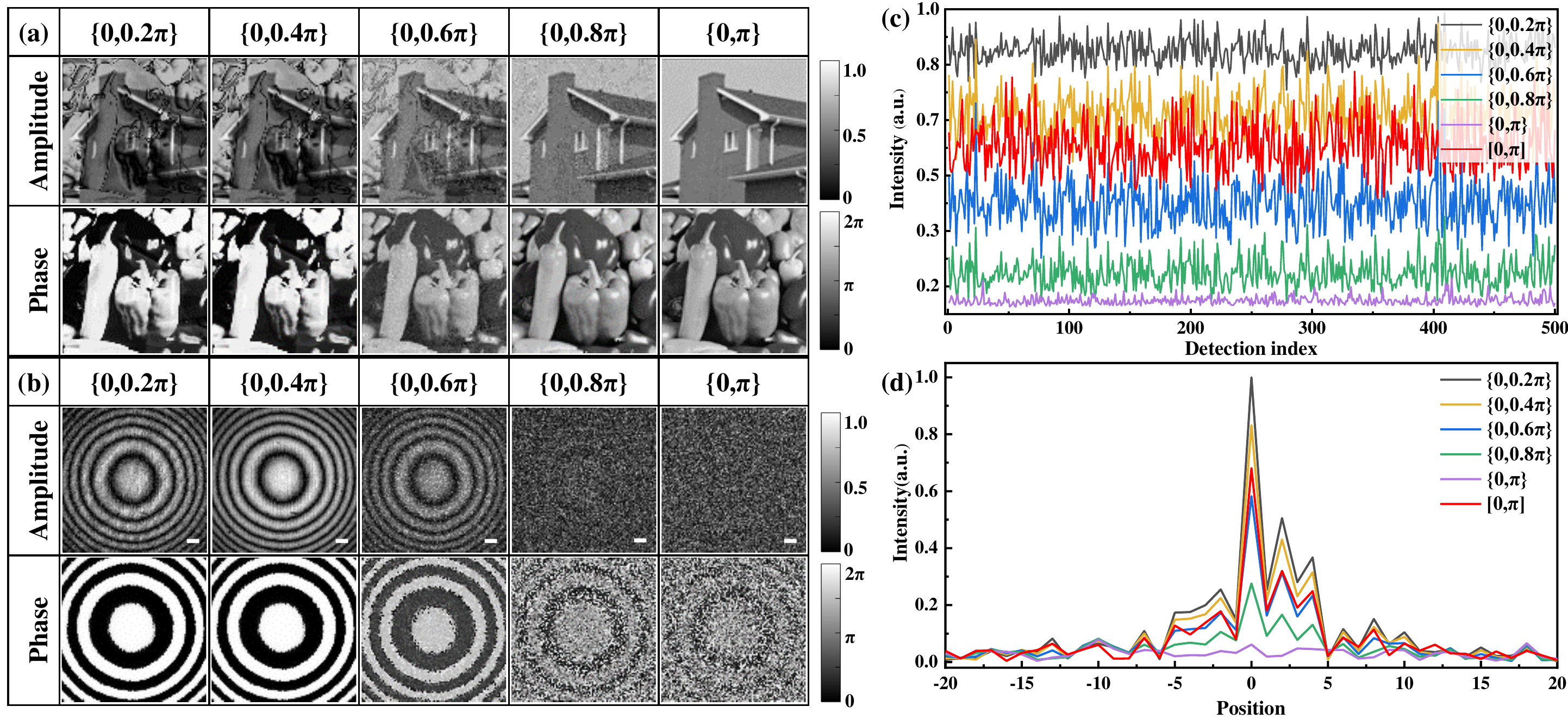}
  \caption{ (a) Simulation reconstructions of a complex object using binary random phase patterns with different phase differences. (b) Experiment reconstructions of an optical lens using binary random phase patterns with different phase differences. (c) A direct comparison of detection intensities when different phase patterns are employed. (d) A slice distribution of 2D Fourier spectrum from a simulation experiment when different phase patterns are employed.}
  \label{fig2}
\end{figure*}

In conventional SPI, illumination modulation is conducted in the amplitude space. For complex objects, however, amplitude-only modulation will cause ambiguity in distinguishing a complex value and its complex conjugate, as both of them have the same detection value as
\begin{equation}
  \begin{aligned}
  y_{i} &=\left |   \bm{a}_{i}^{T}\bm{x}  \right |^{2} =\left |   \bm{a}_{i}^{T}\bm{x}^{*}  \right |^{2}
  \end{aligned}
\end{equation}
Note that $\bm{a}_{i}\in \mathbb{R}^{N }$ is an amplitude-only pattern. An immediate consequence is that Eq.~(\ref{det}) will lead to at least two feasible reconstructions for one object.

For this reason, we use pure phase modulation in our experiment. Phase-only modulation can be done with either binary or grayscale patterns. When we first employ the binary strategy, a crosstalk phenomenon is found between the phase and amplitude reconstructions. Fig.~\ref{fig2}(a) shows the noise-free simulation results under different binary modulation variations.  The complex object has a amplitude distribution of “House” and a phase distribution of “Peppers”. Crosstalk problem is obvious especially when the modulation variation is small.
We can understand this problem as that a small phase variation can not provide sufficient constraint to determine a unique complex-valued solution. Similar phenomenon has also been found in conventional CDI \cite{WilliamsFresnel,ZhangPhase}. 
As we increase the modulation variation, crosstalk decreases, and eventually gets fully removed when $\left \{ 0,\pi \right \}$ patterns are used (maximum phase contrast).
This conclusion can also be verified by experimental data (Fig.~\ref{fig2}(b)). The object is a lens with a pure imaginary-valued distribution function. However, when $\left \{ 0,\pi \right \}$ modulation is employed, despite of the absence of crosstalk, the reconstruction suffers significantly low signal-to-noise ratio (SNR).  Figure.~\ref{fig2}(c) exhibits 500 detection values from each reconstruction shown in Fig.~2(b). Under binary modulation, as the increasing of the phase difference, the detection value trends to become smaller, and the SNR of detected signals become lower. This is because, if we regard the modulation pattern as a binary phase grating, a larger phase contrast will enhance high order diffraction and therefore lower $0$th order intensity. Figure.~\ref{fig2}(d) shows the simulation comparison of the far field spectrum distribution employed in Fig.~\ref{fig2}(a), When $\left \{ 0, \pi\right \}$ binary phase patterns are used, the pinhole detector receives its minimized signal. It results that the reconstruction SNR under this circumstance is extremely low. 

The trade-off between crosstalk and SNR problems indicates that binary pattern is not the best choice. In fact, as we will see, this trade-off can be avoided by expanding the range of phase variation. To guarantee both sufficient modulation constraint and reasonable SNR, we use grayscale patterns that are in a uniform distribution between $0$ and $\pi$. All results in the following sections are achieved with these patterns. A detailed comparison between binary and grayscale modulation also can be found in Fig.~\ref{fig2}(c) and Fig.~\ref{fig2}(d).  Compared with $\left \{ 0, 0.2\pi\right \}$ binary phase pattern, this grayscale modulation retains about 68\% detection intensity of $\left \{ 0, 0.2\pi\right \}$ binary modulation, which guarantees sufficient SNR in the imaging process.

\section{Results and discussions}

\begin{figure}[b!]
  \centering
  \includegraphics[width=0.8\linewidth]{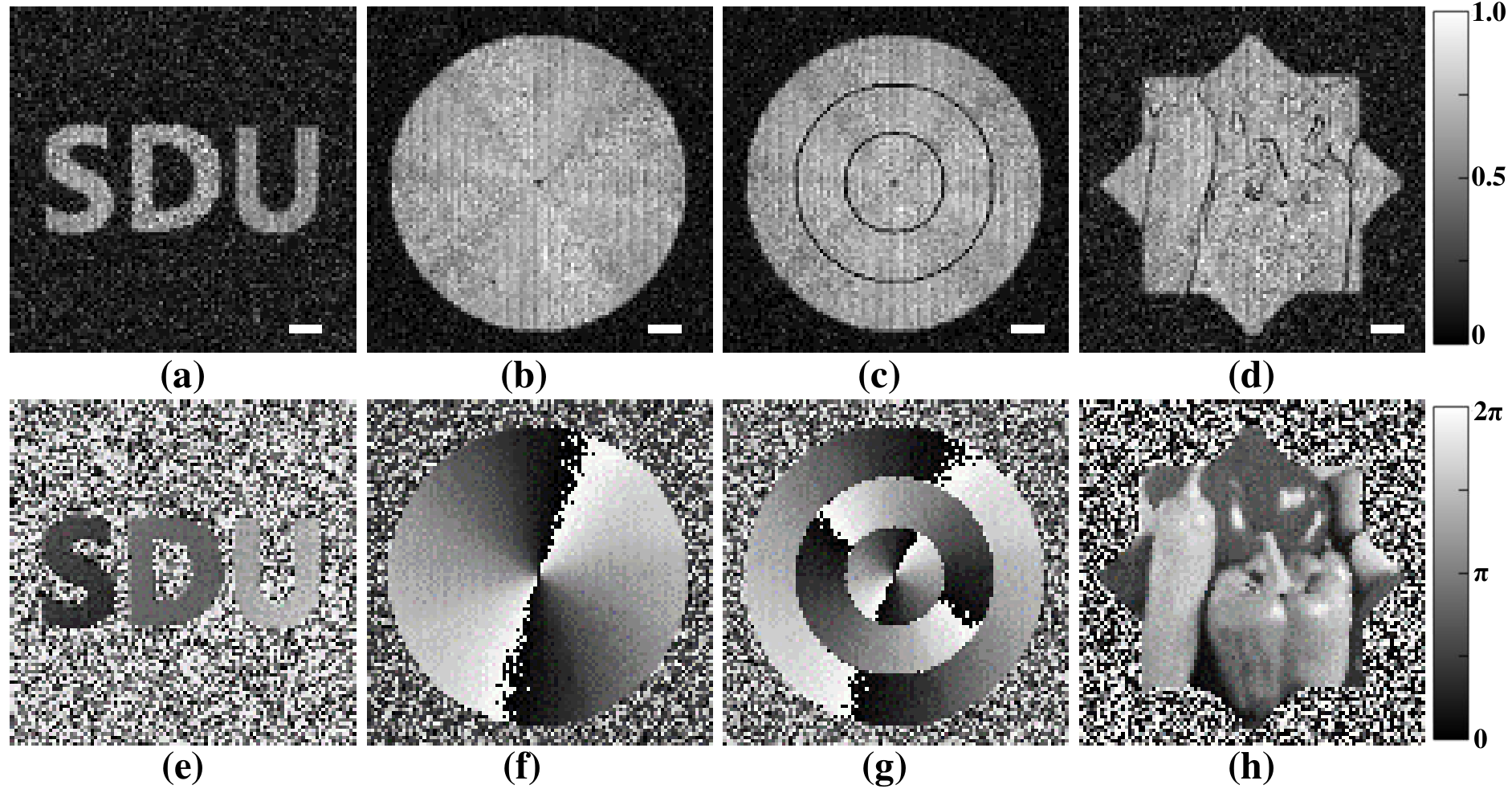}
  \caption{Experiment reconstructions of four SLM-generated complex objects . (a)-(d) The amplitude images of four objects. (e)-(h) Their phase images. The scale bar represents the length of 250$\upmu$m.}
  \label{fig3}
\end{figure}

\begin{figure*}[t!]
  \centering
  \includegraphics[width=\linewidth]{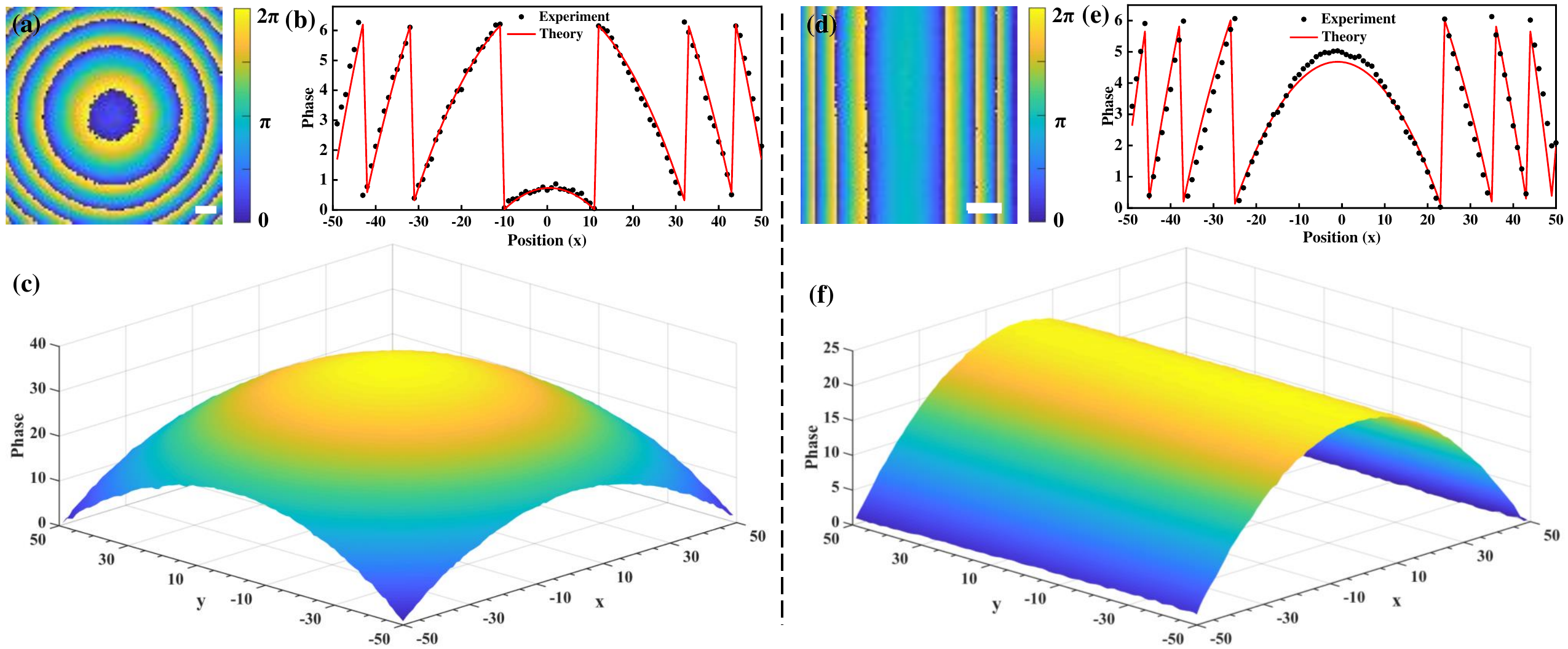}
  \caption{Reconstructions of a plano-convex lens and a cylindrical lens. (a) The phase image of the plano-convex lens with a focal length of 500mm, (b) shows the cross-section of the retrieved phase (black dots) compared with theoretical values (red line). (c) is the 3D display of the unwrapped phase, which is also the surface shape of the lens.  (c) is the phase image of the cylindrical lens with a focal length of 150mm. Again (b) and (c) are the phase cross-section and unwrapped surface. The scale bar represents a length of 250$\upmu$m. }
  \label{fig4}
\end{figure*}

\begin{figure*}[t!]
  \centering
  \includegraphics[width=0.9\linewidth]{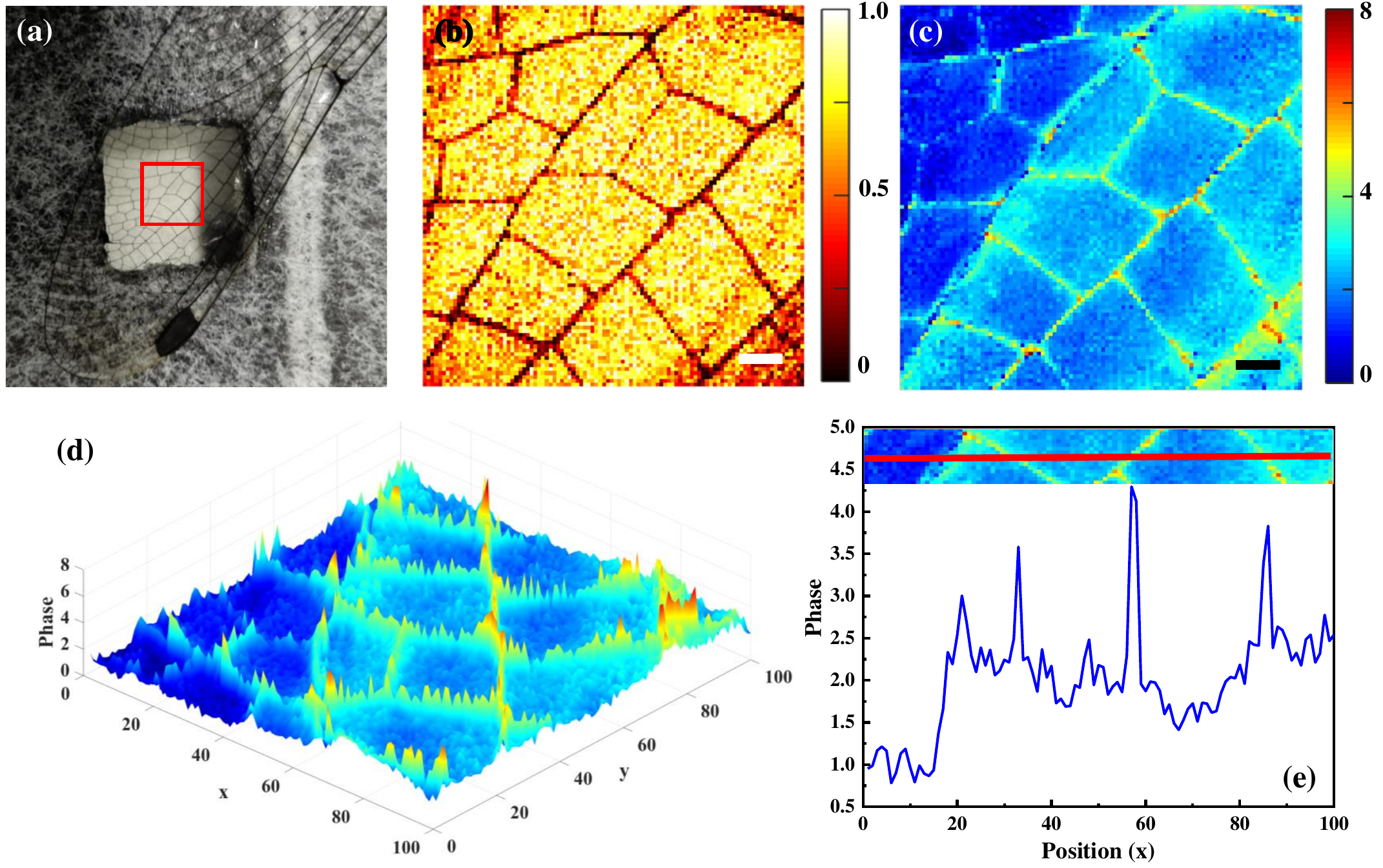}
  \caption{Reconstructions of a dragonfly wing in both amplitude and phase. (a) A camera photo of the wing. The red box indicates the reconstruction region. (b) Amplitude reconstruction. (c) Unwrapped phase reconstruction. (d) Three-dimensional surface of retrieved phase image. The scale bar represents a length of 250$\upmu$m. (e) Cross-section of the phase image shows the sharp changes of phase distribution around the vein nodes.}
  \label{fig5}
\end{figure*}

Our first experiment is done with several SLM generated objects. 
We produce four different complex-wavefront distributions using a SLM (Holoeye Pluto-VIS-016). As shown in Fig.~\ref{fig3}, the first wavefront distribution is three letters 'SDU' with different phase levels ( S: $\pi/2$, D: $\pi$, U: $3\pi/2$). Retrieved amplitude and phase distribution are shown in Fig.~\ref{fig3}(a) and (e). It is important to note that these three letters are isolated to each other, showing that the proposed scheme is feasible for multiple independent objects. Two vortex phase front with different orbital angular momentums are also imaged. Reconstructed amplitude distributions are displayed in Fig.~\ref{fig3}(b)-(c) and the phase distributions are shown in Figure~\ref{fig3}(f)-(g). The results show great consensus with theoretical vortex distribution. The phase singularities of two beams can be clearly distinguished in their amplitude images. Considering a more complicated wavefront distribution,  we generate a complicated field whose amplitude is restricted through an octangle and its phase is modulated by the open source image "Peppers". the reconstruction results are shown in Fig.~\ref{fig3}(d) and (h). Its phase is in good conformity with original 'Peppers' image, and the phase-jump line is also clearly displayed in the amplitude image.

Our second experiment is conducted to measure to optical lenses. The first lens is  a plano-convex lens with a focal length of 500mm, and the second one is a cylindrical lens with a focal length of 150mm. The original reconstructed phase distributions are shown in Fig.~\ref{fig4}(a) and (d). A comparison between the reconstruction phase and the theoretical values are made, corresponding results are shown in Fig.~\ref{fig4}(b) and (e). The correlation coefficients between these experimental and theoretical curves are 0.9196 and 0.9016, respectively. After phase unwrapping, we can get the detected surface distributions, which are shown in Fig.~\ref{fig4}(c) and (f). According to the index of BK7 glass (1.5168), the focal lengths of these two lenses can also be calculated as 511.72mm and 158.72mm. All of these results show that our reconstructions are in a great precision.


In the last experiment, we challenge ourselves with a more ambitious object, a dragonfly wing. The size of the imaging region is set as $2\rm mm \times 2mm$, which is marked as the red box in Fig.~\ref{fig5}(a). The selected region of the wing can be generally divided into two parts, wing membranes and veins. The wing membranes are thin films with high transmittance, while veins are much thicker than the wing membrane and with low transmittance. From the trailing side (top left corner in the imaging region) to the leading side (bottom right corner in the imaging region), the thickness of the wing membrane and vein both gradually increases.  Figure~\ref{fig5}(b) shows its imaged amplitude distribution, both the wing membrane and vein are clearly imaged with a high contrast of brightness. The unwrapped phase distribution is displayed in Fig.~\ref{fig5}(c), and its three-dimensional surface of phase distribution is also shown in Fig.~\ref{fig5}(d). The phase value variation from the top left to the bottom right shows great agreement to the change of thickness distribution of the wing.
The thickness contrast of the wing membranes and veins can also be distinguished clearly. Even the nodes on the veins are very clearly revealed in the phase distribution (the spikes). These extreme sharp phase changes because that, compared to the other parts, the vein nodes are not only much thicker but also denser (higher refraction index), a cross-section of phase image is also plotted in Fig.~\ref{fig5}(e) to show this sharp phase changes. The complex reconstruction of the wing shows great accordance to its biological features.
Especially, the sharp changes in the phase image prove that our scheme is feasible to image complicated complex-valued objects even with large phase variance.

\begin{figure*}[t!]
  \centering
  \includegraphics[width=\linewidth]{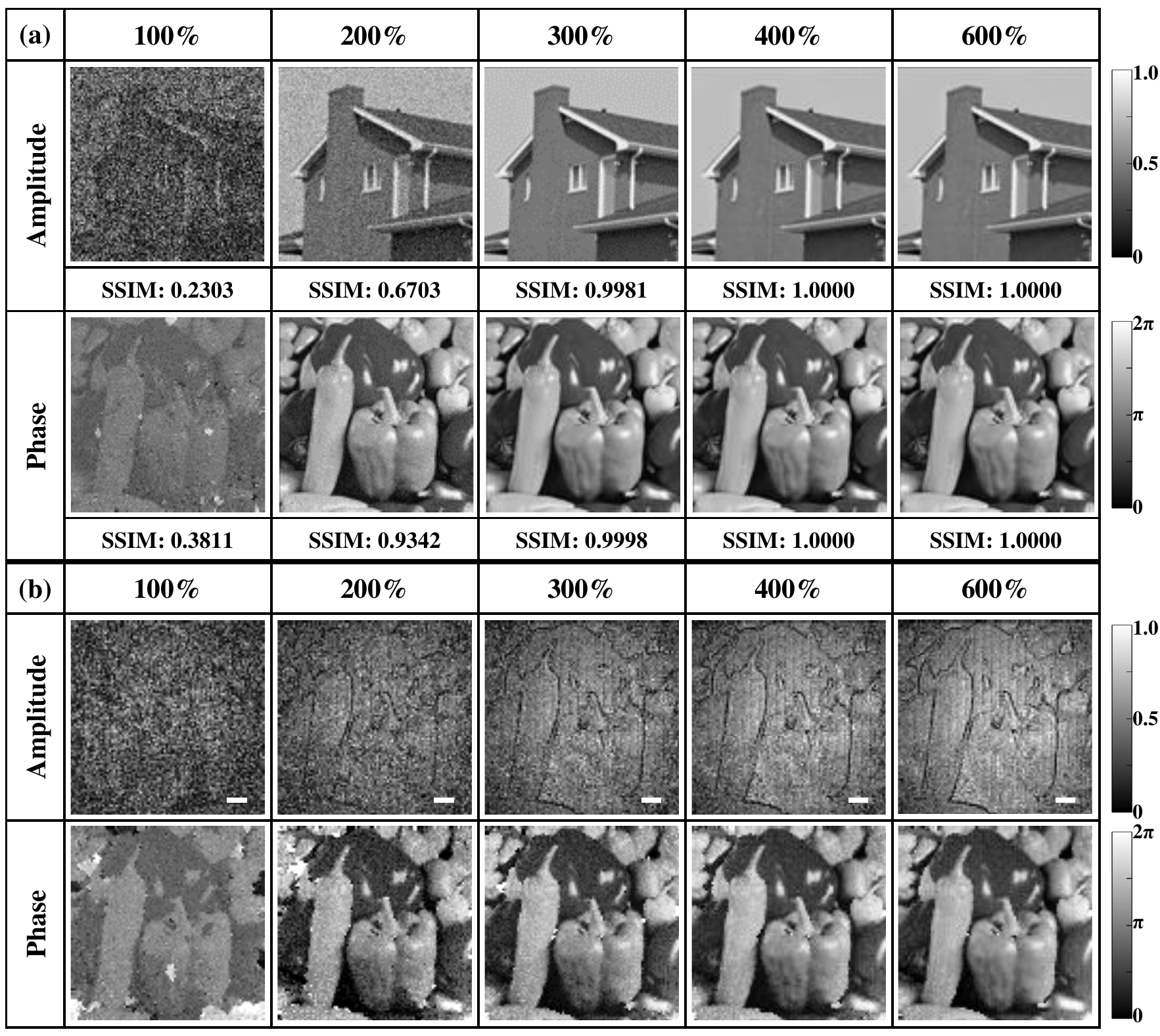}
  \caption{ Reconstructions of a complex object with different sampling ratios. Both (a) simulation and (b) Experimental results are presented. }
  \label{fig6}
\end{figure*}


All reconstructions above are done with a sampling ratio of 600\% to maximize imaging quality. Nevertheless, as described in the theory section, in phase retrieval problem, normally a four times sampling ratio is required to determine an unique solution of an unknown complex field. In practical, thanks for the possible correlated information between adjacent pixels, or other internal constraints such as nonnegativity of the amplitude, it is possible to obtain sufficient image information below this bound. In Fig. 6 we investigate the sampling ratio requirement in our approach in both simulation and experiment. In Fig.~\ref{fig6}(a), we exhibit the simulation results under different sampling ratios from 100\% to 600\%. The object is a complex field with a amplitude distribution of “House” and a phase distribution of “Peppers”. Quantitative evaluations using structural similarity index (SSIM) are marked under each reconstruction. As shown in the results, although a perfect reconstruction happens at a sampling ratio of 400\%, the reconstruction is already “nearly perfect” at a sampling ratio of 300\%. Fig.~\ref{fig6}(b) shows experimental results with noise involved. The SLM-generated object has a uniform amplitude, while its phase is in a distribution of “Peppers”. It shows that, although the reconstruction still a slight promotion visually when the sampling ratio increases from 400\% to 600\%, the result image is already in a reasonable quality with 400\% samples, or even with the 300\% samples. This is a clear proof that our scheme is sampling-efficient.

\section{conclusion}

To summarize, we have developed a diffraction imaging technique for complex objects via single-pixel camera. Our single-pixel diffraction imaging (SPDI) scheme retrieves complex images by performing phase-modulated illumination and far-field single-pixel detection. The structured illumination not only functions for spatial sampling for two-dimensional imaging, but also provides extra constraints in phase retrieval. Reconstruction is achieved by using TAF algorithm to minimize a non-convex quadratic loss function. By introducing sampling redundancy, our SPDI requires no a priori information of the objects. Both simulations and experiments have shown that our SPDI scheme can image objects even with extremely rough phase distribution. Moreover, single-pixel detection has reduced the requirement for detector resolution. Therefore, it can be applied to those wavelengths that focal plane detector array technology is technically or financially expensive. On the other side, the pinhole in front of the detector actually plays a role for spatial mode selection. It indicates that this SPDI scheme could be achieved with broadband light source. Based on these advantages, we believe that our SPDI technology has universal applications from long-wavelength, e.g. microwave and terahertz light, to short-wavelength, e.g. X-rays, electrons and neutrons. At last, it should be noticed that our work is done under the assumption of thin film objects, e.g., objects are taken as two dimensional. Recently, several works have extended CDI into a tomography scheme to reconstruct 3D complex field\cite{tomography,Coupled}. In future work, effort should be devoted to extended our SPDI scheme for 3D complex field imaging.

\section*{Declaration of Competing Interest}
The authors declare that they have no known competing financial interests or personal relationships that could have appeared to influence the work reported in this paper.

\section*{CRediT authorship contribution statement}
\textbf{Xianye Li}: Conceptualization, Validation, Data curation, Methodology, Writing - original draft. \textbf{Yafei Sun}: Methodology,Validation. \textbf{Yikang He}: Validation. \textbf{Xun Li}: Writing - Review \& Editing. \textbf{Baoqing Sun}: Writing - Review \& Editing, Funding acquisition, Supervision.

\section*{Acknowledgments}
This work was supported by the National Natural Science Foundation of China (NSFC) (61675117), Shandong Joint Funds of Natural Science (ZR2019LLZ003-1) and Shandong University Inter-discipline Research Grant.

\bibliography{main}

\begin{thebibliography}{10}
\expandafter\ifx\csname url\endcsname\relax
  \def\url#1{\texttt{#1}}\fi
\expandafter\ifx\csname urlprefix\endcsname\relax\def\urlprefix{URL }\fi
\expandafter\ifx\csname href\endcsname\relax
  \def\href#1#2{#2} \def\path#1{#1}\fi

\bibitem{QPIreview}
K.~Lee, K.~Kim, J.~Jung, J.~Heo, S.~Cho, S.~Lee, G.~Chang, Y.~Jo, H.~Park,
  Y.~Park, Quantitative phase imaging techniques for the study of cell
  pathophysiology: from principles to applications, Sensors 13~(4) (2013)
  4170--4191.

\bibitem{TIE}
K.~Nugent, T.~Gureyev, D.~Cookson, D.~Paganin, Z.~Barnea, Quantitative phase
  imaging using hard x rays, Physical Review Letters 77~(14) (1996) 2961.

\bibitem{phasecontrast}
S.~Wilkins, T.~E. Gureyev, D.~Gao, A.~Pogany, A.~Stevenson, Phase-contrast
  imaging using polychromatic hard x-rays, Nature 384~(6607) (1996) 335--338.

\bibitem{abbe}
E.~Abbe, Archiv f{\"u}r mikroskopische, Anatomie 9 (1873) 413--418.

\bibitem{zernike1955discovered}
F.~Zernike, How i discovered phase contrast, Science 121~(3141) (1955)
  345--349.

\bibitem{Digitalholography}
R.~Collier, Optical holography, Elsevier, Amsterdam, 2013.

\bibitem{Phase-shifting}
I.~Yamaguchi, T.~Zhang, Phase-shifting digital holography, Optics Letters
  22~(16) (1997) 1268--1270.

\bibitem{wavefrontsensing}
U.~Schnars, C.~Falldorf, J.~Watson, W.~J{\"u}ptner, Digital holography and
  wavefront sensing., Springer, Berlin, 2016.

\bibitem{QPIBio}
Y.~Park, C.~Depeursinge, G.~Popescu, Quantitative phase imaging in biomedicine,
  Nature Photonics 12~(10) (2018) 578--589.

\bibitem{holo3D}
G.~Nehmetallah, P.~P. Banerjee, Applications of digital and analog holography
  in three-dimensional imaging, Advances in Optics and Photonics 4~(4) (2012)
  472--553.

\bibitem{wolf}
E.~Wolf, Solution of the phase problem in the theory of structure determination
  of crystals from x-ray diffraction experiments, Physical Review Letters
  103~(7) (2009) 075501.

\bibitem{phasereview}
Y.~Shechtman, Y.~C. Eldar, O.~Cohen, H.~N. Chapman, J.~Miao, M.~Segev, Phase
  retrieval with application to optical imaging: a contemporary overview, IEEE
  Signal Processing Magazine 32~(3) (2015) 87--109.

\bibitem{cdixray}
I.~Robinson, R.~Harder, Coherent x-ray diffraction imaging of strain at the
  nanoscale, Nature Materials 8~(4) (2009) 291--298.

\bibitem{electron}
C.~B. Carter, D.~B. Williams, Transmission electron microscopy: Diffraction,
  imaging, and spectrometry, Springer, Berlin, 2016, Ch.~3, p.~81.

\bibitem{neutron}
R.~Hill, C.~Howard, Quantitative phase analysis from neutron powder diffraction
  data using the rietveld method, Journal of Applied Crystallography 20~(6)
  (1987) 467--474.

\bibitem{GS}
Z.~Zalevsky, D.~Mendlovic, R.~G. Dorsch, Gerchberg--saxton algorithm applied in
  the fractional fourier or the fresnel domain, Optics Letters 21~(12) (1996)
  842--844.

\bibitem{fienup}
J.~R. Fienup, Reconstruction of an object from the modulus of its fourier
  transform, Optics Letters 3~(1) (1978) 27--29.

\bibitem{nphard}
P.~M. Pardalos, S.~A. Vavasis, Quadratic programming with one negative
  eigenvalue is np-hard, Journal of Global Optimization 1~(1) (1991) 15--22.

\bibitem{support}
J.~R. Fienup, Reconstruction of a complex-valued object from the modulus of its
  fourier transform using a support constraint, JOSA A 4~(1) (1987) 118--123.

\bibitem{Fourier}
R.~N. Bracewell, The Fourier transform and its applications, Vol. 31999, New
  york: Mcgraw-hill, New York, 1986, Ch.~12, p. 241.

\bibitem{CMI1}
F.~Zhang, J.~Rodenburg, Phase retrieval based on wave-front relay and
  modulation, Physical Review B 82~(12) (2010) 121104.

\bibitem{CMI2}
F.~Zhang, B.~Chen, G.~R. Morrison, J.~Vila-Comamala, M.~Guizar-Sicairos, I.~K.
  Robinson, Phase retrieval by coherent modulation imaging, Nature
  communications 7 (2016) 13367.

\bibitem{SPIreview}
M.~P. Edgar, G.~M. Gibson, M.~J. Padgett, Principles and prospects for
  single-pixel imaging, Nature Photonics 13~(1) (2019) 13--20.

\bibitem{abedi2020single}
M.~Abedi, B.~Sun, Z.~Zheng, Single-pixel compressive imaging based on random
  dog filtering, Signal Processing 178 (2020) 107746.

\bibitem{SPICS}
N.~Radwell, K.~J. Mitchell, G.~M. Gibson, M.~P. Edgar, R.~Bowman, M.~J.
  Padgett, Single-pixel infrared and visible microscope, Optica 1~(5) (2014)
  285--289.

\bibitem{csholo}
P.~Clemente, V.~Dur{\'a}n, E.~Tajahuerce, P.~Andr{\'e}s, V.~Climent, J.~Lancis,
  Compressive holography with a single-pixel detector, Optics letters 38~(14)
  (2013) 2524--2527.

\bibitem{liu2019complex}
R.~Liu, S.~Zhao, P.~Zhang, H.~Gao, F.~Li, Complex wavefront reconstruction with
  single-pixel detector, Applied Physics Letters 114~(16) (2019) 161901.

\bibitem{liu2018single}
Y.~Liu, J.~Suo, Y.~Zhang, Q.~Dai, Single-pixel phase and fluorescence
  microscope, Optics express 26~(25) (2018) 32451--32462.

\bibitem{hu2019single}
X.~Hu, H.~Zhang, Q.~Zhao, P.~Yu, Y.~Li, L.~Gong, Single-pixel phase imaging by
  fourier spectrum sampling, Applied Physics Letters 114~(5) (2019) 051102.

\bibitem{phasecodecpr}
R.~Pedarsani, D.~Yin, K.~Lee, K.~Ramchandran, Phasecode: Fast and efficient
  compressive phase retrieval based on sparse-graph codes, IEEE Transactions on
  Information Theory 63~(6) (2017) 3663--3691.

\bibitem{coded}
E.~J. Candes, X.~Li, M.~Soltanolkotabi, Phase retrieval from coded diffraction
  patterns, Applied and Computational Harmonic Analysis 39~(2) (2015) 277--299.

\bibitem{oversampling}
B.~Leshem, R.~Xu, Y.~Dallal, J.~Miao, B.~Nadler, D.~Oron, N.~Dudovich, O.~Raz,
  Direct single-shot phase retrieval from the diffraction pattern of separated
  objects, Nature Communications 7~(1) (2016) 1--6.

\bibitem{SDP}
I.~Waldspurger, A.~d’Aspremont, S.~Mallat, Phase recovery, maxcut and complex
  semidefinite programming, Mathematical Programming 149~(1) (2015) 47--81.

\bibitem{phasemax}
T.~Goldstein, C.~Studer, Phasemax: Convex phase retrieval via basis pursuit,
  IEEE Transactions on Information Theory 64~(4) (2018) 2675--2689.

\bibitem{TAF}
G.~Wang, G.~B. Giannakis, Y.~C. Eldar, Solving systems of random quadratic
  equations via truncated amplitude flow, IEEE Transactions on Information
  Theory 64~(2) (2017) 773--794.

\bibitem{Wirtingerflow}
E.~J. Candes, X.~Li, M.~Soltanolkotabi, Phase retrieval via wirtinger flow:
  Theory and algorithms, IEEE Transactions on Information Theory 61~(4) (2015)
  1985--2007.

\bibitem{signalrec}
R.~Balan, P.~Casazza, D.~Edidin, On signal reconstruction without phase,
  Applied and Computational Harmonic Analysis 20~(3) (2006) 345--356.

\bibitem{ratio}
A.~Conca, D.~Edidin, M.~Hering, C.~Vinzant, An algebraic characterization of
  injectivity in phase retrieval, Applied and Computational Harmonic Analysis
  38~(2) (2015) 346--356.

\bibitem{WilliamsFresnel}
G.~J. Williams, H.~M. Quiney, B.~B. Dhal, C.~Q. Tran, K.~A. Nugent, A.~G.
  Peele, D.~Paterson, M.~D. de~Jonge, Fresnel coherent diffractive imaging,
  Physical Review Letters 97~(2)  025506.

\bibitem{ZhangPhase}
F.~Zhang, G.~Pedrini, W.~Osten, Phase retrieval of arbitrary complex-valued
  fields through aperture-plane modulation, Physical Review A 75~(4)  043805.

\bibitem{tomography}
R.~Horstmeyer, J.~Chung, X.~Ou, G.~Zheng, C.~Yang, Diffraction tomography with
  fourier ptychography, Optica 3~(8) (2016) 827.

\bibitem{Coupled}
M.~Kahnt, J.~Becher, D.~Brückner, Y.~Fam, C.~G. Schroer, Coupled ptychography
  and tomography algorithm improves reconstruction of experimental data, Optica
  6~(10) (2019) 1282.

\end{thebibliography}

\end{document}